\documentstyle[12pt,epsf]{article}
\begin{document}
\newcommand{\lsim}{\mathrel{\lower4pt\hbox{$\sim$}}
\hskip-12.5pt\raise1.6pt\hbox{$<$}\;}

\newcommand{\gsim}{\mathrel{\lower4pt\hbox{$\sim$}}
\hskip-12.5pt\raise1.6pt\hbox{$>$}\;}

\def\jpsi{J/\psi~}
\def\BAR{\overline}
\def\xba{\overline}
\def\fm{{\cal M}}
\def\fl{{\cal L}}
\def\ufs{\Upsilon(5S)}
\def\ufour{\Upsilon(4S)}
\def\xcp{X_{CP}}
\def\ynotcp{Y}
\vspace*{-.5in}
\def\etap{\eta^\prime}
\def\rhobar{\overline\rho}
\def\etabar{\overline\eta}

\def\uglu{\hskip 0pt plus 1fil
minus 1fil} \def\uglux{\hskip 0pt plus .75fil minus .75fil}

\def\slashed#1{\setbox200=\hbox{$ #1 $}
    \hbox{\box200 \hskip -\wd200 \hbox to \wd200 {\uglu $/$ \uglux}}}
\def\slpar{\slashed\partial}
\def\sla{\slashed a}
\def\slb{\slashed b}
\def\slc{\slashed c}
\def\sld{\slashed d}
\def\sle{\slashed e}
\def\slf{\slashed f}
\def\slg{\slashed g}
\def\slh{\slashed h}
\def\sli{\slashed i}
\def\slj{\slashed j}
\def\slk{\slashed k}
\def\sll{\slashed l}
\def\slm{\slashed m}
\def\sln{\slashed n}
\def\slo{\slashed o}
\def\slp{\slashed p}
\def\slq{\slashed q}
\def\slr{\slashed r}
\def\sls{\slashed s}
\def\slt{\slashed t}
\def\slu{\slashed u}
\def\slv{\slashed v}
\def\slw{\slashed w}
\def\slx{\slashed x}
\def\sly{\slashed y}
\def\slz{\slashed z}

\rightline{AMES-HET-01-02}
\rightline{BNL-HET-01/9}

\begin{center}

{\large\bf
CP Asymmetry Measurements in \boldmath$\jpsi K^0$ and the CKM Paradigm
}

\vspace{.2in}

David Atwood$^{1}$\\
\noindent Department of Physics and Astronomy, Iowa State University, Ames,
IA\ \ \hspace*{6pt}50011\\
\medskip

Amarjit Soni$^{2}$\\
\noindent Theory Group, Brookhaven National Laboratory, Upton, NY\ \
11973\\
\footnotetext[1]{email: atwood@iastate.edu\hskip1.5in $^2$email:
soni@bnl.gov}
\end{center}
\vspace{.15in}

\begin{quote}
{\bf Abstract:}

Recent experimental observations of CP asymmetry in $B \to \jpsi + K^0$
constitute the first significant signal of CP violation outside the
neutral kaon system; thus they represent an important milestone to test
the CKM paradigm. We, therefore, undertake a critical appraisal of the
existing experimental and theoretical inputs used to deduce constraints on
$\sin 2 \beta$ and other important parameters and thus find, in
particular, $\sin2 \beta > 0.47$ at 95\% CL which is completely compatible
with the combined experimental result: $\sin2 \beta = 0.48 \pm 0.16$,
representing an important success of the CKM model of CP violation.
Searches for new physics in $B$ decays to $\jpsi + K^0$ like final states
will require improved precision;  we make some suggestions to facilitate
these. We also present a global fit including the new CP asymmetry
measurements in $B\to \jpsi+K^0$ as an additional input yielding e.g.\
$\gamma =(29^\circ\to 67^\circ)$, 
$\etabar = (.20\to.37)$, 
$\rhobar=(.08 \to .36)$, 
$J_{CP}=(1.8\to3.1)\times10^{-5}$,
$Br(K^+\to\pi^+\nu\bar\nu)=(0.52\to 0.92)\times 10^{-10} $, 
$Br(K_L\to\pi^0\nu\bar\nu)= (0.11\to 0.31)\times 10^{-10}$ and $\Delta
m_{B_s} = (15.2\mbox{--}28.3)~ps^{-1}$, all at 95\%
CL\null.

\end{quote}

\newpage


Recent measurements of CP asymmetry in the $B^0\to \jpsi K^0$ 
and related processes~\cite{psik_note}
by the
BELLE \cite{belle_ref} and BaBar~\cite{babar_ref} detectors at the KEK and
SLAC $B$-factories together with the earlier measurement by
CDF~\cite{cdf_ref}
constitute the first significant  signal of CP violation outside of the
neutral kaon system. As such, they afford a unique test of the CKM
paradigm~\cite{cabibo,kob_mask} along with possible clues for the presence
of new physics. In this work, we
present a critical appraisal of the theoretical and experimental inputs
used in such analysis and suggest  directions for
improvement to facilitate more precise studies of the  future.
Furthermore, we specifically address the question: When should one
begin to take seriously a conflict between the value of
$\sin2\beta(\equiv\sin2\phi_1 [7])$ as measured in $B^0\to J/\psi K^0$
decays and that deduced from other processes?

In the Kobayashi-Maskawa~\cite{kob_mask} description of CP violation which
is an integral part of the Standard Model (SM), a ``clean'', (i.e.\ free of
hadronic uncertainties) determination of the angle $\beta$ of the
unitarity triangle~\cite{u_triang} can be made by measuring the time
dependent partial rate asymmetry for $B_d$ and $\xba B_d$ decays to a
common final state which is a CP eigenstate~\cite{carter}:

\begin{eqnarray}
a(t)={
\Gamma(B_d(t)\to f)-\Gamma(\xba B_d(t)\to f)
\over
\Gamma(B_d(t)\to f)+\Gamma(\xba B_d(t)\to f)
}
=
-f_{CP} \sin2\beta \sin(\Delta m_d t) \label{eqn1}
\end{eqnarray}

\noindent
where $f_{CP}$ is the CP eigenvalue of $f$
and
$\Delta m_d $
is the  $B^0_d$-$\xba B^0_d$ mass difference.

The recent
experimental determination of the phase $\beta$
by the BELLE~\cite{belle_ref}, BaBar~\cite{babar_ref} and
CDF~\cite{cdf_ref} groups, 
through the use of eqn.~(\ref{eqn1})
is:

\begin{eqnarray}
\sin2\beta=
\left \{
\begin{array}{ll}
0.58~{+0.32\atop -0.34}~{+0.09\atop -0.10} &   ~~~~~[BELLE]\\
0.34~\pm 0.20 \pm 0.05               &       ~~~~~[BaBar]\\
0.79~\pm 0.44                         &       ~~~~~[CDF]
\end{array}
\right .
\label{eqn2}
\end{eqnarray}

\noindent
By combining these with the earlier results \cite{consist_note} which
are consistent with these but have larger errors, we get the average

\begin{eqnarray}
\sin2\beta = 0.48\pm 0.16
\label{w_average} \label{eqn3}
\end{eqnarray}

%
%
%
%

In the Wolfenstein representation \cite{wolf}
of the CKM matrix, two of the
parameters are rather well known \cite{PDB}, $\lambda= 0.2196\pm 0.0023$,
$A=0.85\pm.04$. 
Considerable efforts are under way for a better
determination
of the remaining two parameters  $\rho$ and $\eta$, or,
equivalently \cite{buras},
$\rhobar=\rho(1-\lambda^2/2)$
and
$\etabar=\eta(1-\lambda^2/2)$ which are introduced in the
generalization which maintains unitarity to higher order in
$\lambda$ \cite{twelve,schmidtler}.

Currently, there are four important inputs that constrain the values of
$\rhobar$ and $\etabar$ \cite{plasz}:
(1)~The $\epsilon_K$ parameter of indirect CP-violation of the $K^0\xba
K^0$ system, (2)~$R_{uc}\equiv |V_{ub}/V_{cb}|$ which is deduced from the
fraction of semi-leptonic $B$-decays to charmless final states,
(3)~$\Delta m_d$, the mass difference
which drives
$B_d\xba B_d$ mixing and
(4)~the LEP bound on $\Delta m_s$ ($\geq \hbar~15~ps^{-1}$), the mass
difference
which drives
$B_s\xba B_s$ mixing.

In order to determine the parameters $\rhobar$ and $\etabar$ and the
sensitivity of physical quantities to theoretical inputs, we will
use a set of  ``nominal'' inputs for evaluating the four physical
quantities mentioned above. In our nominal input we take
$R_{uc}=0.085\pm 0.017$, although later we will
also consider the case where the error is inflated to 0.0255 [i.e.\ to
a total of 30\%] for a more
``conservative''
interpretation of the current   results.

The SM expression for $\epsilon_K$ in terms of CKM elements
involves the non-perturbative hadronic parameter
$B_K\equiv <K|(\xba s \gamma_\mu(1-\gamma_5)d)^2|\xba
K>/(8/3f_K^2m_K^2)$.
For the corresponding  renormalization group invariant (to
NLO) quantity we take $\hat B_K=0.86\pm 0.06\pm 0.14$ in our nominal input
from lattice calculations \cite{lellouch}. The lattice calculation of
this quantity, with the
stated error is quite safe as it has now been calculated with staggered
fermions as well as with  the newer method of domain wall quarks  which
have  superior realization of chiral symmetry. The error due to
unquenching and  due to SU(3) breaking are each estimated at around
$\sim5\%$ and
are included in the stated
systematic error \cite{lellouch}.

For the evaluation of $\Delta m_d$ we need $f_{B_d}\sqrt{\hat B_{B_d}}$
and the analysis of $\Delta m_s$ requires the $SU(3)$ breaking ratio
$\xi$ $\equiv$  $f_{B_s} \sqrt{\hat B_{B_s}}/ (f_{B_d}$\break $\sqrt{\hat
B_{B_d}})$. These quantities have been studied extensively using lattice
Monte Carlo methods. Bernard in his review at Lattice 2000 \cite{bernard}
(as well as Aoki \cite{aoki}) gave their values as $f_{B_d}\sqrt{\hat
B_{b_d}} = 230\pm 40$ MeV and $\xi=1.16\pm0.05$. 
We have inflated the error 
on 
$f_{B_d}\sqrt{\hat B_{B_d}}$ 
by about 20\%
to 50 MeV for our nominal
input. Also, in our nominal input we will increase the error on $\xi$ to
0.08 to reflect our concern that the ``direct'' evaluation of the SU(3)
breaking via:


\begin{eqnarray}
{
<B_s| [\xba b \gamma_\mu (1-\gamma_5) s]^2 | \xba B_s>
\over
<B_d| [\xba b \gamma_\mu (1-\gamma_5) d]^2 | \xba B_d>
}
=
{
m^2_{s}
\over
m^2_{d}
}
\xi^2
\label{eqn4}
\end{eqnarray}

\noindent tends to give somewhat larger central values \cite{bernardtwo}
although with rather large errors so that within errors the results are
compatible in the two methods. Since ${\Delta m_s\over \Delta m_d}$ is an
extremely important phenomenological constant it will certainly be very
helpful if improved lattice studies via (\ref{eqn4}) re-confirm the value
of $\xi$ obtained via the $ f_{B_s}\sqrt{B_{B_s}}/ (f_{B_d}\sqrt{B_{B_d}})
$ method \cite{bernard}.

In Table~1 we collect the ``nominal'' inputs for the important parameters
taken from experiment and from the lattice. The Table also shows our
inputs for the four parameters [$R_{uc}$, $\hat B_K$, $f_{B_d}\sqrt{\hat
B_{B_d}}$ and $\xi$] for the ``conservative'' solution. For this case we
take $\hat B_K=0.90\pm0.06\pm0.14$ as there are preliminary indications
from some lattice calculations of a few percent increase of this quantity
in dynamical simulations \cite{kilcup}. Similarly there are indications
\cite{gimenez} that $B_{B_d}$ tends to decrease by about 11\%; therefore,
in our conservative choice we will take $f_{B_d}\sqrt{\hat B_{B_d}}=217\pm
50$ MeV and also we will inflate the error on $\xi$ to 0.1. Note that
these ``conservative'' choices are made as they tend to lower the value of
$(\sin2\beta)_{min}$.


Fig.~1(a) shows the resulting constraints from our nominal input in the
$\etabar$-$\rhobar$ plane\cite{da_alg}. 
Fig.~1(b) shows the bounds on $\sin2\beta$,
e.g.\ 
$0.51\leq\sin2\beta\leq0.88$ at 95\% CL. 
In the second part of Table~1 we show the resultant confidence intervals
for a number of quantities that depend on the CKM matrix.  Based on the
``nominal'' and ``conservative'' inputs, we quote the angular quantities
$\sin 2\beta$, $\sin 2\alpha$ and $\gamma$ as well as the parameters
$\rhobar$ and $\etabar$ and the ratio $|V_{td}/V_{ts}|$.  We also include
the Jarlskog invariant of the CKM matrix, 
$J_{CP}\equiv A^2\lambda^6\etabar$
~\cite{buras,jarlskog}, and give the corresponding predictions for the
kaon decays $K^+\to\pi^+\nu\bar\nu$ and
$K_L\to\pi^0\nu\bar\nu$. These two decay modes are sensitive to the
magnitude and imaginary part of $V_{td}$ respectively.

%
%
%
%
%
%
%
%
%
%
%
%

We have also studied the sensitivity of the results on $\hat B_K$,
$f_{B_d}\sqrt{\hat B_{B_d}}$, $\xi$, $R_{uc}$ and on the $\Delta m_s$
bound;  of these the dependence on $\Delta m_s$ is especially interesting
and important. As Fig.~2 shows the bound on $\sin2\beta$ is quite
insensitive to the experimental bound on $\Delta m_s$ so long as $\Delta
m_s\gsim 9\hbar~ps^{-1}$. Lower values of this quantity do tend to effect
the bound. It is somewhat reassuring that so long as the LEP bound
\cite{lepb,slcb} is $\geq 9\hbar~ps^{-1}$, i.e. almost 50\% of its stated
value, it does not have a major effect on $\sin2\beta$.

%
%
%
%
%

%
%
%

Fig.~3 summarizes 
our results on the
bounds on
$\sin2\beta$ with various inputs. 
In addition to the two sets of
inputs mentioned above, i.e.\ ``nominal''  and
``conservative'', we also include here the following 2
cases.

\begin{description}

\item [\rm ~~~1.] Do not use $f_{B_d}\sqrt{\hat B_d}$ and use only
$\epsilon_K$, $R_{uc}$, 
$\frac{\Delta m_s}{\Delta m_d}$ with the ``conservative'' input from
Table~1.

\item [\rm ~~~2.] Disregard $\frac{\Delta m_s}{\Delta m_d}$ 
as it may be of some concern that while, at present,
we only have a bound on ${\Delta m_s}$, 
used in conjunction with the parameter $\xi$ which 
the lattice gives with rather tight errors, it ends up playing a rather
important role; thus use only  $\epsilon_K$,
$R_{uc}$ and $\Delta m_d$.

\end{description}

%
%


We see that $\sin2\beta\gsim.47$  (at 95\% CL)
if all four of the experimental inputs are
used.  Fig.~3 also shows the experimental results for $\sin2\beta$. It is
clear that the experimental information available so far is completely
compatible with theoretical expectations of the CKM model; there is no
glaring need of new physics or new CP-odd phase(s). Searches for the
effects of new CP-odd phase(s)  \cite{newphys}, at least in these
channels, will require precision studies.

An interesting feature of Fig.~3 is that 
the $\sin 2\beta$ interval depends little
on whether or not we include $f_{B_d}\sqrt{B_{B_d}}$ 
(i.e. the input $\Delta m_d$).
This is 
because the main effect of this constraint is on
the magnitude of $V_{td}$ while $\beta$ 
depends on its phase. Thus, $\beta$ and
$f_{B_d}\sqrt{B_{B_d}}$ tend not to be correlated.
Another 
notable aspect
of Fig.~3 is  that the lower bound
on $\sin2\beta$ gets appreciably 
reduced to $\sim0.35$ 
(95\% CL)
if $\Delta m_s$ is not included as an input. Thus,
if improved experimental determinations of $\sin2\beta$ end up finding a
value lower than in eqn~(\ref{eqn3}) then that may be indicative of
problems with the $\Delta m_s$ bound of $15~\hbar~ps^{-1}$.

It is useful to compare our analysis to the one by Gilman {\it et al}
\cite{plasz} given in the Particle Data Book. There are small
differences with us in various inputs such as $\xi$ and $\Delta
m_{B_s}$;  the most notable one is that our central value for
$f_{B_d}\sqrt{\hat B_d}$ is bigger by about 10\%. This is in line with
recent assessment of effects of quenching \cite{bernard}. The main
effect of these differences is that our value for $\rhobar$ tends to be
somewhat bigger compared to Gilman {\it et al}. \cite{global}.

Since a very important source of error in  the hadronic parameters 
that we took from the lattice originates from the quenched approximation,
improvements are quite challenging and are likely to take considerable
time and effort. Therefore, the bounds that we deduced on $\sin2\beta$ and
other quantities are not likely to change any time soon.  On the other
hard, experiments from $e^+e^-$ and hadronic $B$-facilities are expected
to rapidly increase the pool of data by almost an order of magnitude or
even more in the next year or two. Therefore, our bound should be able to
 facilitate tests of the CKM paradigm more definitively in the near
future.

%
%

The case when $\epsilon_K$ is not used in the input data set is of special
significance as then the remaining three inputs ($R_{uc}$, $\Delta m_d$,
$\frac{\Delta m_s}{\Delta m_d}$) are all from {\it CP-conserving\/} $B$
experiments [Fig.~4]. The resulting constraint on $\etabar$ is therefore
especially important as a non-vanishing lower bound on $\etabar$ would
then imply the need for the CP violation phase even when accounting for CP
conserving $B$ experiments~\cite{etasign_note}.
Furthermore, if the CKM model is correct then
the lower bound on $\etabar$ thus deduced from $B$ experiments must
satisfy the constraint from $\epsilon_K$ on the $\etabar$-$\rhobar$
plane; in particular $\epsilon_K$ requires $\etabar>.12$ (see Fig.~1(a)).
Unfortunately the current accuracy of input data sets is not sufficient to
give a 95\% CL nonvanishing lower bound on $\etabar$ although at 68\% CL
that is the case: 
$0.084 <\etabar<0.324$ 
at 68\% CL and
$0.014<\etabar<0.410$ 
at 95\% CL\null.  Thus as the accuracy on
$\frac{V_{ub}}{V_{cb}}$, $\Delta m_s$ and 
the hadronic parameters from the lattice  improves it is
very likely that in the near future the CP
conserving $B$ experiments would lead to a non-vanishing lower bound on
$\etabar$ even at 95\% CL thereby providing a completely non-trivial test of
the CKM model. 
Meantime, the $\sin2\beta$ measurements of eq.~(\ref{eqn3}) via
$B^0\to\jpsi K^0$ may also now be included with the other three
$B$-experiments [$R_{uc}$, $\Delta m_d$ and $\frac{\Delta m_s}{\Delta
m_d}$] to give a solution entirely from $B$-physics. One thus finds,
$\etabar= 0.13\to 0.27 $ at 68\% CL and 
$\etabar=0.07\to 0.35$ at 95\% CL.

Since CP asymmetry results, eq.~(\ref{eqn3}), on $B^0\to\jpsi K^0$ are
found to be completely compatible with the expectations of the CKM
model, we can now include these latest results, so that our input set
now consists of $\epsilon_K$, $R_{uc}$, $\Delta m_d$, $\Delta m_s$ and
$\sin2\beta$ to obtain the new global fits (see Fig.~5). The
corresponding 68\% and 95\% CL results for various CKM parameters and
other quantities of interest are given in Table~2.

In conclusion, from Table 1 and Fig. 3 it is apparent that
if $\sin2 \beta$ is eventually measured to be significantly
$\lsim 0.47$, the CKM model of CP violation, or some
subset of the theoretical inputs regarding matrix elements or
the $\Delta m_s$ bound, will have a problem. 
On the other hand, if $\sin2 \beta \gsim 0.47$ improvements in the
theoretical inputs and or additional experimental input(s)
will be needed to refine our tests of the Standard Model
CKM picture. Bearing that in mind we now briefly mention a few 
areas where experimental
and/or theoretical progress could be very useful.

\begin{enumerate}

\item Along the same lines as mentioned above,  it would be
extremely useful to deduce the entire 
unitarity triangle from $K$-physics so that one can make more decisive
comparisons, in particular of the Jarlskog invariant~\cite{jarlskog},
$J_{CP}$, deduced from $B$-physics with that from $K$-physics and not just
contend ourselves with the lower bound on $\etabar$ that $\epsilon_K$
gives or the constraints from $\epsilon_K$ only. Studies of
$K^+\to\pi^+\nu\xba\nu$ can give a rather clean determination of 
$|V_{td}|$ and study of $K^0_L\to\pi^0\nu\xba\nu$ can give $\etabar$ or
$J_{CP}$ very cleanly and can be very useful in that regard
\cite{burastwo}.

\item Although direct experimental measurement of $f_B$ via $B^\pm\to
\tau^\pm +\nu$ still remains difficult due to the low branching ratio, some
experimental information on $f_B$ could be obtained via the radiative
decays $B^+\to \ell^++\nu_\ell +\gamma$. The branching ratio is expected
to be \cite{atwood} $\sim5\times10^{-6}$ using $f_B=200$ MeV and
$\frac{V_{ub}}{V_{uc}}=0.085$. Therefore, by adding $e^+$, $e^-$, $\mu^+$,
$\mu^-$ the effective branching ratio is about $2\times10^{-5}$ and may
well already be accessible. Furthermore, the photon spectrum is relatively
hard as it originates from a spin-flip; the initial pseudoscalar meson
emits the photon and becomes a vector or axial vector state for
annihilation into $\ell+\nu$ without having to pay the penalty of helicity
suppression. The calculations are model dependent but perhaps an accuracy
of $\sim30\%$ on $f_B$ could be attained through this method and that
may provide a useful experimental check on the lattice calculations.

\item It is clearly important to determine $\Delta m_s$ via $B_s$-$\xba
B_s$ oscillation. This is likely to come from CDF/$D\emptyset$ at the
Tevatron and perhaps also from HERAB, TeVB, and LHCB\null. Meantime, it
would be  useful to attain information on $\frac{V_{td}}{V_{ts}}$ via
$B^0\to\rho^0+\gamma$ \cite{a_b_s}. The ratio $\frac{B^0 \to
\rho^0+\gamma}{B^\to 
K^\ast+\gamma}$ should give 
$\frac{|V_{td}|}{|V_{ts}|}$ up to SU(3) corrections to a very good
approximation. Calculation of $B\to 
\rho(K^\ast)+\gamma$ involves only one form factor \cite{bernardthree}. It
would be extremely useful to use lattice and other methods to calculate
the SU(3) breaking effects on that form factor \cite{formfactor} which
would be needed to relate measurements of $(B^0\to\rho^0+\gamma)/(B\to
K^\ast+\gamma)$ to $|V_{td}/V_{ts}|$.

\end{enumerate}

\bigskip

We acknowledge useful discussions with Sinya Aoki, Adam Falk and Fred
Gilman. This research was supported in part by US DOE Contract Nos.\
DE-FG02-94ER40817 (ISU) and DE-AC02-98CH10886 (BNL)

\newpage

%
%
%
%
%

\newcommand{\tstrut}{\vline height11pt depth7pt width0pt}
\begin{table}[h]
\begin{center}
\caption{Fits using ``nominal'' and ``conservative'' values for the
four input parameters.
The QCD correction coefficients $\eta_1$, $\eta_2$, $\eta_3$ and $\eta_b$
are taken
from~\cite{eta_ref} and $V_{cb}=0.040\pm 0.002$~\cite{errnote}.
\label{tabone}}
\hspace*{-.8in}{\footnotesize\begin{tabular}{|l|c|c|c|c|}
\hline
\tstrut{ Input Quantity} & \multicolumn{2}{|c|}{Nominal} &
\multicolumn{2}{|c|}{Conservative} \\
\hline
& \multicolumn{2}{|c|}{} & \multicolumn{2}{|c|}{} \\
\tstrut$R_{uc}\equiv|V_{ub}/V_{cb}|$ & \multicolumn{2}{|c|}{$0.085\pm.017$} &
\multicolumn{2}{|c|}{$0.085\pm .0255$} \\
\tstrut$f_{B_d}\sqrt{\hat B_{B_d}}$ & \multicolumn{2}{|c|}{$230\pm 50$ MeV}
&
\multicolumn{2}{|c|}{$217 \pm 50$ MeV} \\
\tstrut$\xi$ & \multicolumn{2}{|c|}{$1.16\pm0.08$} &
\multicolumn{2}{|c|}{$1.16\pm0.10$} \\
\tstrut$\hat B_K$ & \multicolumn{2}{|c|}{$0.86\pm0.15$} &
\multicolumn{2}{|c|}{$0.90\pm0.15$} \\
& \multicolumn{2}{|c|}{} & \multicolumn{2}{|c|}{} \\
\hline
\tstrut Output Quantity & 68\% CL & 95\% CL & 68\% CL & 95\% CL \\
\cline{1-5}
\tstrut$\sin2\beta$ & 
$0.60\to 0.80$ & $0.51 \to 0.88$ & 
$0.58\to 0.83$ & $0.47 \to 0.93$ \\
\tstrut$\sin2\alpha$ & 
$-0.81\to-0.18$ & $-0.96\to  0.17$ &
$-0.82\to-0.14$ & $-0.96\to  0.27$ \\
\tstrut$\gamma$ & 
$37.1^\circ\to 55.3^\circ$ & 
$30.2^\circ\to 65.4^\circ$
&
$36.4^\circ\to 56.3^\circ$ & 
$29.5^\circ\to 63.3^\circ$ \\
\tstrut$\etabar$ & 
$0.25 \to 0.35$  &  $0.21 \to 0.41$ & 
$0.24 \to 0.36$  &  $0.20 \to 0.44$ \\
\tstrut$\rhobar$ & 
$0.17 \to 0.32$ & $0.10 \to 0.39$ & 
$0.16 \to 0.34$ & $0.07 \to 0.42$ \\
\tstrut$|V_{td}/V_{ts}|$ & 
$0.17 \to 0.20$ & $0.15 \to 0.21$ & 
$0.16 \to 0.20$ & $0.15 \to 0.22$
\\
$\Delta m_{B_s}~(ps^{-1} )$  
& 
$16.3 \to 23.4$ &   $15.3 \to 29.3$ &
$16.3 \to 24.6$ &   $15.3 \to 31.8$
\\
\tstrut$J_{CP}$ & 
$(2.2\to 2.9)\times10^{-5}$ & 
$(1.9\to 3.4)\times10^{-5}$ &
$(2.1\to 3.0)\times 10^{-5}$ & 
$(1.8\to 3.5)\times10^{-5}$ \\
\tstrut$Br(K^+\to\pi^+\nu\xba\nu)$ & 
$(0.57\to 0.77)\times10^{-10}$ &
$(0.49\to 0.90)\times10^{-10}$ & 
$(0.55\to 0.78)\times10^{-10}$ &
$(0.47\to 0.91)\times10^{-10}$ \\
\tstrut$Br(K_L\to\pi^0\nu\xba\nu)$ & 
$(0.16\to 0.29)\times10^{-10}$ &
$(0.12\to 0.38)\times10^{-10}$ & 
$(0.15\to 0.30)\times10^{-10}$ &
$(0.11\to 0.41)\times10^{-10}$ \\
\hline
\end{tabular}}
\end{center}
\end{table}
%
%
%
%
%
%
%
%
%
%

%
%

\newpage

\begin{table}[h]
\begin{center}
\caption{Fits using the measured value of $\sin2\beta$ (see
eq.~(\ref{eqn3})) plus the four inputs with ``nominal'' values given in
Table~1.\label{tabtwo}} 
\bigskip
{\begin{tabular}{|l|c|c|}
\hline
\tstrut Output Quantity & 68\% CL & 95\% CL  \\
\hline
\tstrut$\sin2\beta$ & 
$0.56 \to 0.73$ & $0.49 \to 0.81$  
\\
\tstrut$\sin2\alpha$ & 
$-0.87 \to -0.24$ & 
$-0.98 \to 0.16$  
\\
\tstrut$\gamma$ & 
$36.5^\circ \to 56.2^\circ$ & 
$29.4^\circ \to 67.3^\circ$
\\
\tstrut$\etabar$ & 
$0.23 \to 0.32$ & 
$0.20 \to 0.37$  
\\
\tstrut$\rhobar$ & 
$0.15 \to 0.30$ & 
$0.08 \to 0.36$  
\\
\tstrut$|V_{td}/V_{ts}|$ & 
$0.17 \to 0.20$ & 
$0.15 \to 0.22$ 
\\
$\Delta m_{B_s}~(ps^{-1})$ & 
$16.1 \to 22.8$  & 
$15.2 \to 28.3$ 
\\
\tstrut$J_{CP}$ & 
$(2.1\to 2.7)\times10^{-5}$ & 
$(1.8\to 3.1)\times10^{-5}$  
\\
\tstrut$Br(K^+\to\pi^+\nu\xba\nu)$ & 
$(0.60\to 0.81)\times10^{-10}$ &
$(0.52\to 0.92)\times10^{-10}$  
\\
\tstrut$Br(K_L\to\pi^0\nu\xba\nu)$ & 
$(0.14\to 0.24)\times10^{-10}$ &
$(0.11\to 0.31)\times10^{-10}$  
\\
\hline
\end{tabular}}
\end{center}
\end{table}

%
%
%
%

\newpage

\begin{center}
{\bf Figure Captions}
\end{center}

{\bf Figure 1:}

(a)~A plot of the allowed region in the $\rhobar$-$\etabar$ plane using
our nominal inputs.  
The 1-$\sigma$ constraints resulting from
$\epsilon_K$ are shown with the dashed lines;  
those from 
$|V_{ub}/V_{cb}|$ 
are shown with the dash-dotted lines and the
1-$\sigma$ constraints resulting from the $B_d\xba B_d$ oscillation rate
are shown with the dotted lines. The thin solid line, obtained by using
the 95\% CL bound $\Delta m_s>15 ps^{-1}$ and $\xi\le1.32$ (which is
the 2-$\sigma$ bound), indicates the limit
from $B_s\xba B_s$ oscillation.  The solid coutours indicate the 
interval
allowed by the combined constraints
with the inner contour indicating the $68\%$ confidence interval
and the outer one  the $95\%$ confidence interval.

(b)~The likelihood function for various values of $\sin 2\beta$. The
dotted box indicates the 68\% confidence interval while the dashed box
indicates the 95\% confidence interval.

\medskip

{\bf Figure 2:}
The solid lines indicate the upper and lower confidence (95\% CL)
interval in $\sin 2\beta$ as a function of the bound on $\Delta 
m_s$ in units of $\hbar ps^{-1}$.

\medskip

{\bf Figure 3:} The 68\% and 95\% CL allowed intervals in $\sin 2\beta$ are
shown for various  fits to the CKM matrix discussed in the text. Also
shown are  the $1-\sigma$
ranges  for the data from BaBar, BELLE , CDF as well as their
combined result.

\medskip

{\bf Figure 4:}
The $\etabar$-$\rhobar$ plot as in Figure~1(a) (i.e.\ with nominal
input) except that the
$\epsilon_K$ data are not used in the global fit shown so that the input
set consists of only the CP conserving B-experiments, namely 
$R_{uc}$,
$\Delta m_d$ and $\Delta m_s/\Delta m_d$; 
in this figure we take the sign
of $\etabar$ to be positive~\cite{etasign_note}.

%
%
%
%

\medskip

{\bf Figure 5:}
Figure shows the global fit using the nominal four inputs
from Table 1 but in addition including also the $\sin2\beta$
measurements (see eq.~(\ref{eqn3})).


\newpage

~
\begin{figure}
\epsfxsize 5 in
\mbox{\epsfbox{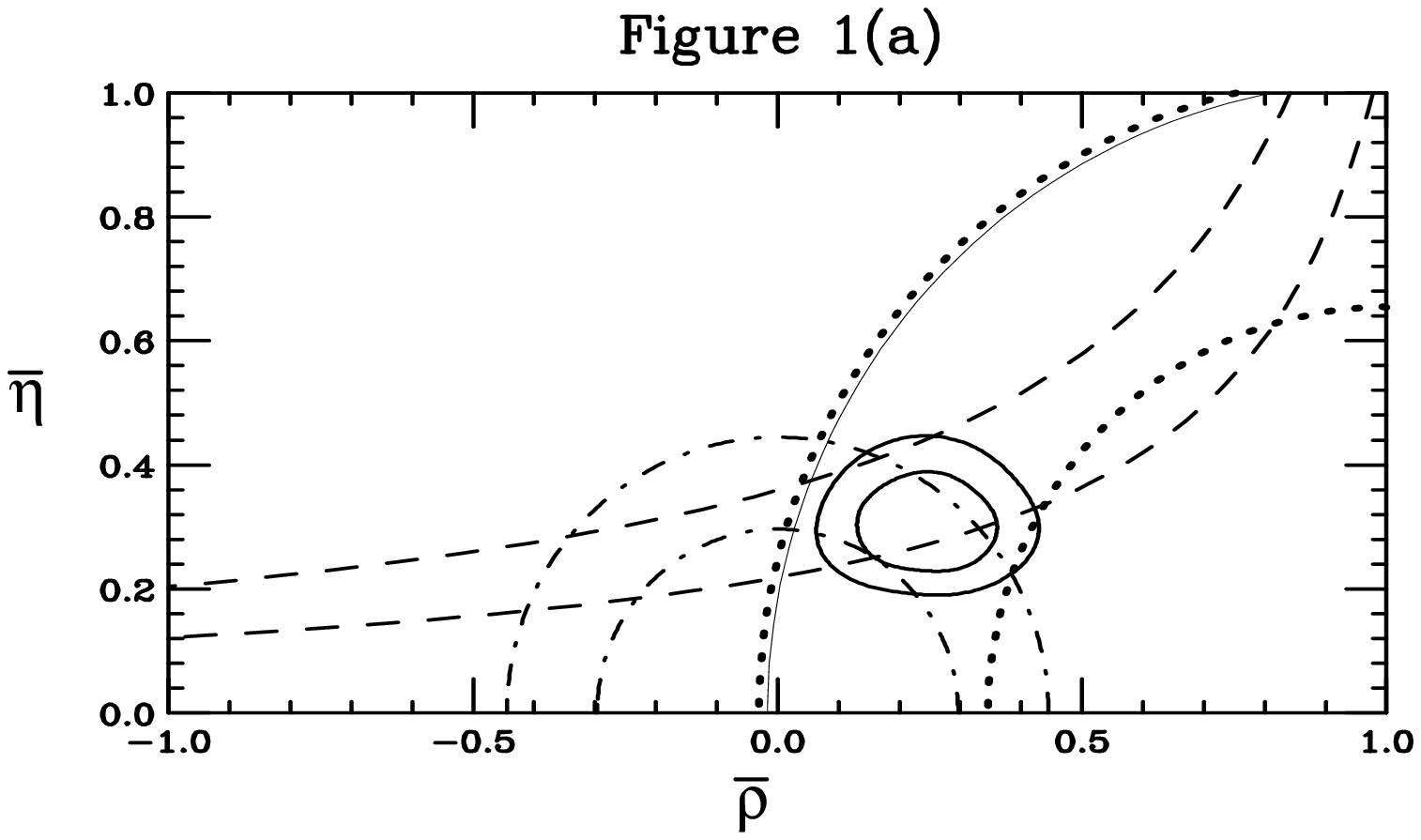}}
\end{figure}

\newpage

~
\begin{figure}
\epsfxsize 5 in
\mbox{\epsfbox{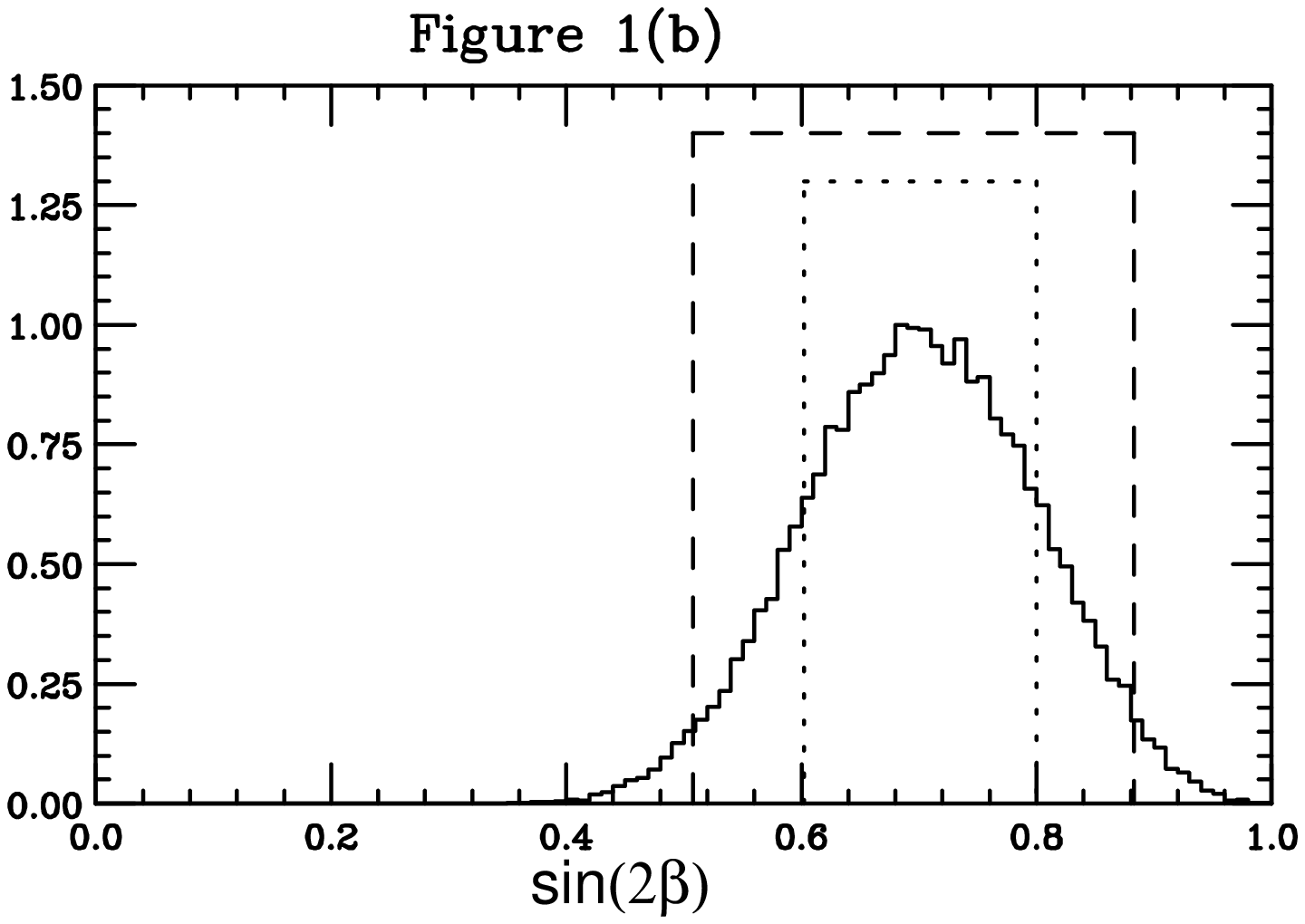}}
\end{figure}

\newpage

~
\begin{figure}
\epsfxsize 5 in
\mbox{\epsfbox{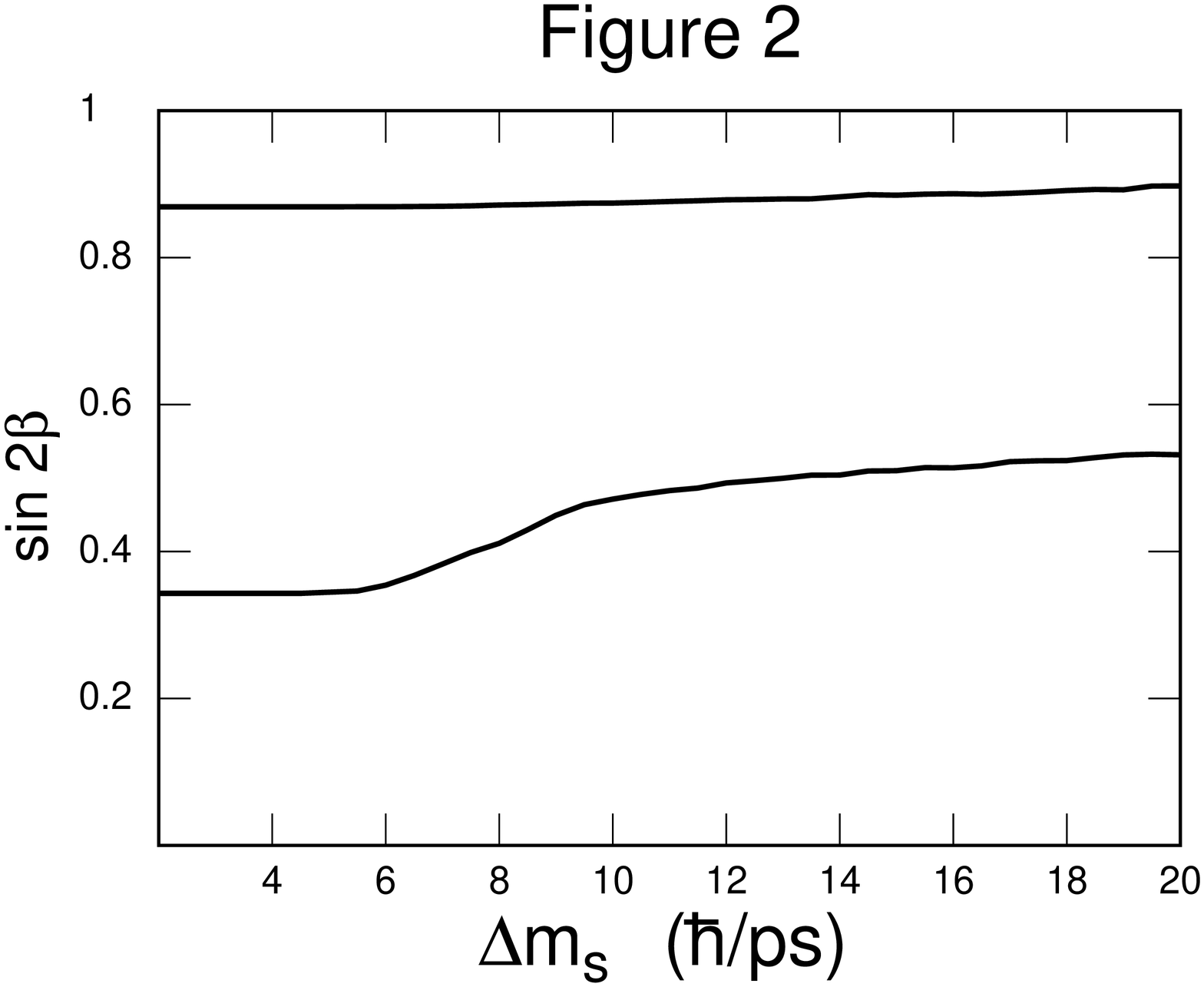}}
\end{figure}

\newpage
~
\begin{figure}
\epsfxsize 6 in
\mbox{\epsfbox{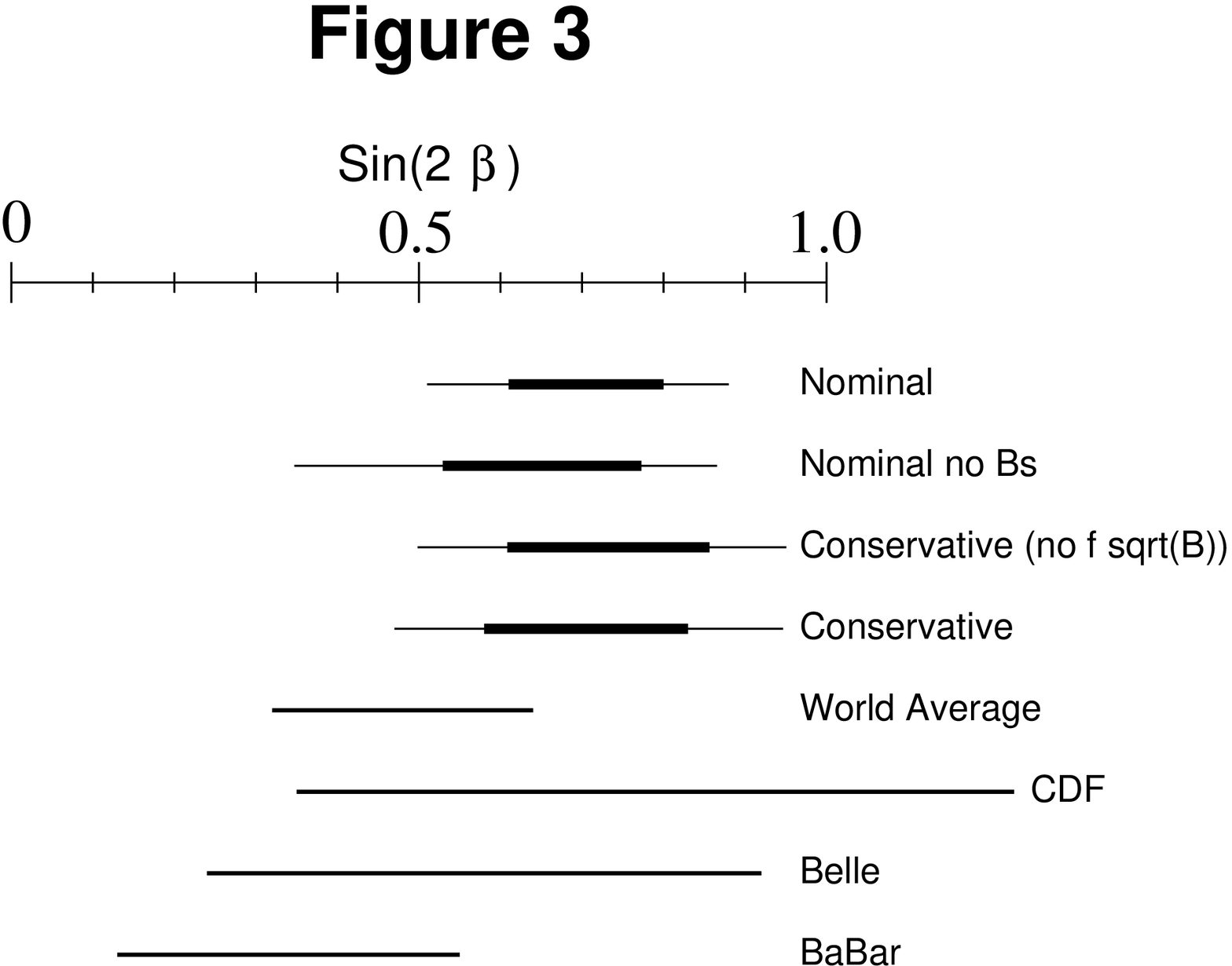}}
\end{figure}

\newpage
~
\begin{figure}
\epsfxsize 5 in
\mbox{\epsfbox{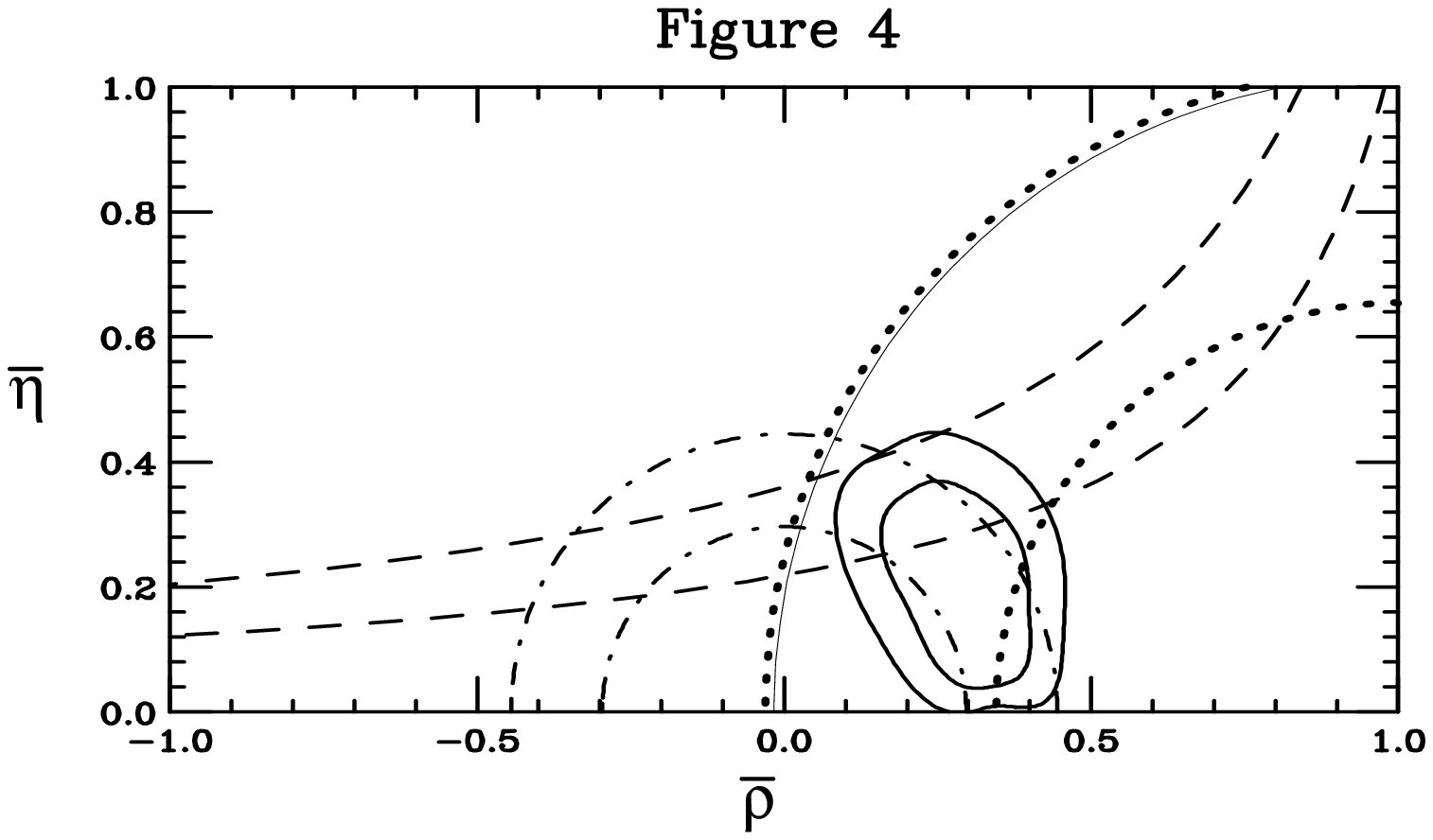}}
\end{figure}

%
%
%
%

\newpage
~
\begin{figure}
\epsfxsize 5 in
\mbox{\epsfbox{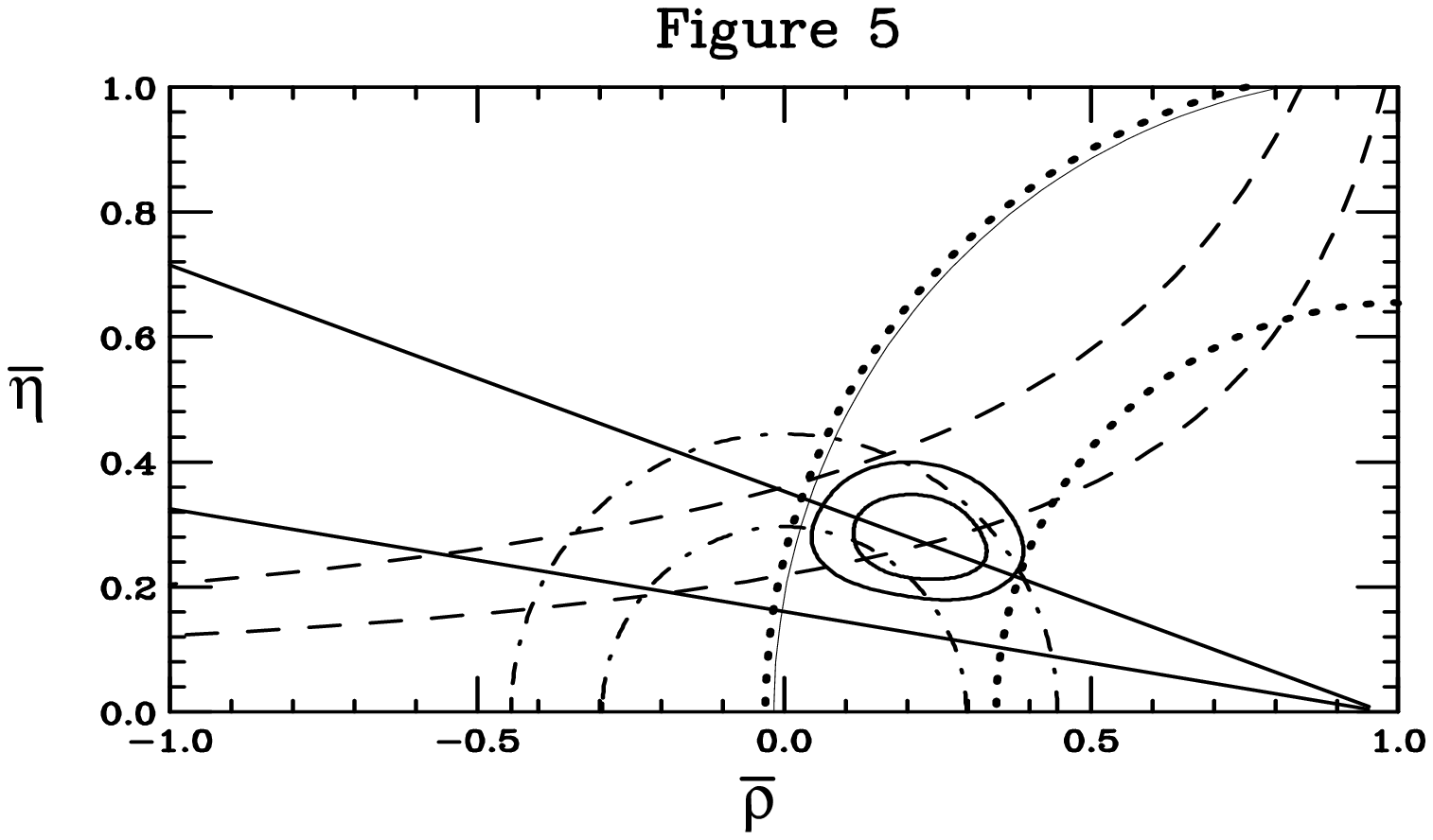}}
\end{figure}


\begin{thebibliography}{99}
\bibitem{psik_note}
Some of the related processes measured are $\psi^\prime K^0$, $\eta_c
K^0$,$\dots$. For simplicity we will collectively call these $\jpsi K^0$.
\bibitem{dualnote} Note the dual notation $(\alpha,\beta,\gamma) \equiv
(\phi_2,\phi_1,\phi_3)$ in use at present; however, for definiteness,
here we choose to use $\alpha,\beta,\gamma$.   
\bibitem{da_alg}
In order to fit the available data, we have interpreted the
experimental and theoretical inputs stated above as Gaussian probability
distributions. We then used a Monte Carlo algorithm to select a number
(40000) of sample sets of parameters with a probability given by combined
probability of these distributions. From these parameter sets, we then
extracted the confidence intervals of $\sin 2\beta$, the allowed region of
the $\etabar-\rhobar$ plot and other parameters of interest.  
\bibitem{global} Given that considerable theoretical input is used in a
global analysis such as ours, the strategy of Gaussian errors that we
have adopted \cite{da_alg} may be of some concern and indeed, for this
reason, some groups have opted for the so called scanning method. We
have two comments to make in this regard. Firstly, the theoretical
input used in here comes exclusively from lattice calculations and the
relevant studies by now are rather matured having been examined 
extensively by a
variety of techniques. We, therefore, believe that the theoretical
errors being quoted are quite reliable. Secondly, we have checked that
when the same input is used our method tends to give constraints quite
similar to the scanning method; we explicitly checked this with the
first paper in Ref.~16.  
\bibitem{etasign_note}
Of course the sign of
$\etabar$ could not be determined without CP violation, only the fact
that the magnitude of $\etabar$ was non-zero.   
\bibitem{formfactor} Strictly speaking to evaluate the absolute rate,
the form factor at $q^2=0$ is needed which may be difficult for the
lattice. However, for study of the SU(3) breaking 
ratio, the dependence on $q^2$ is likely to be minimal allowing for an
accurate extrapolation to the physical point, $q^2=0$, from data
collected at small values of $q^2$.   
\bibitem{errnote} An increase in the error on $V_{cb}$ by 50\% (i.e.\
to $\pm.003$) has little effect (i.e. $\lsim10\%$) on the quantities
given in the output here.   
\end{thebibliography}
\end{document}